\documentclass[letter]{aa}
\usepackage{graphicx}
\usepackage{txfonts}

\usepackage{color}
\usepackage{array}
\usepackage{caption}
\usepackage{subcaption}
\usepackage{amsmath,amssymb,amsfonts}
\usepackage{gensymb, footnote}
\usepackage{sidecap}
\usepackage{changepage}
\usepackage{tabu}
\usepackage[toc,page]{appendix}
\usepackage{multirow}
\usepackage[]{hyperref}
\hypersetup{
   colorlinks = true,
   citecolor ={cyan}
}

\usepackage{titlesec}

\usepackage{tablefootnote}
\usepackage{threeparttable}
\usepackage{upgreek}

\usepackage{stfloats}
\usepackage{placeins}

\title{MWC 656}
\author{Soetkin Janssens}
\date{April 2023}

\defcitealias{Casares_et_al_2014}{CNR14}
\defcitealias{Rivinius_et_al_2022}{RKC22}

\begin{document}
   \title{MWC 656 is unlikely to contain a black hole} \titlerunning{}

   \author{S. Janssens \inst{1} %\thanks{\email{soetkin.janssens@kuleuven.be}}
           \and T. Shenar\inst{2}
           \and N. Degenaar\inst{2}
           \and J. Bodensteiner\inst{3}
           \and H. Sana\inst{1}
           \and J. Audenaert\inst{1}
           \and A. J. Frost\inst{4}
%           \and S. Simón-Díaz\inst{4}
          }

   \institute{Institute of Astronomy, KU Leuven, Celestijnenlaan 200D, 3001 Leuven, Belgium\\ \email{soetkin.janssens@kuleuven.be}
    \and
    Anton Pannekoek Institute for Astronomy, Science Park 904, 1098 XH, Amsterdam, The Netherlands    
    \and
    European Organisation for Astronomical Research in the Southern Hemisphere (ESO), Karl-Schwarzschild-Str 2, D-85748, Garching, Germany
    \and European Southern Observatory, Alonso de Cordova 3107, Vitacura, Casilla 19001, Santiago de Chile, Chile
%    \and 
%    Instituto de Astrofísica de Canarias, 38 200 La Laguna, Tenerife, Spain
          }    

   \date{}

% \abstract{}{}{}{}{} 
% 5 {} token are mandatory
 
  \abstract
   {MWC 656 was reported as the first known Be star with a black-hole (BH) companion in a 60\,d period. The mass of the proposed BH companion is estimated to be between $4-7 M_{\odot}$. This estimate is based on radial velocity (RV) measurements derived from the Fe\,{\sc ii} $\lambda$4583 emission line of the Be star disc and from the He\,{\sc ii} $\lambda$4686 emission line, assumed to be formed in a disc around the putative BH.
   }
   {Using new high-resolution spectroscopic data, we investigate whether MWC 656 truly contains a BH.
   }
   {We used the cross-correlation method to calculate the RVs of both the Be star and the He\,{\sc ii} $\lambda$4686 emission line and we derive a new orbital solution. We also performed disentangling to look for the spectral signature of a companion.
   }
   {
   We derive an orbital period of $59.028\pm 0.011$\,d and a mass ratio $q = M_{\text{He {\sc ii}}}/M_{\text{Be}} = 0.12 \pm 0.03$, much lower than the previously reported $q = 0.41 \pm 0.07$. Adopting a mass of the Be star of $M_{\text{Be}} = 7.8\pm2.0 M_{\odot}$, the companion has a mass of $0.94 \pm 0.34 M_{\odot}$. For the upper limit of $M_{\text{Be}} = 16M_{\odot}$ and $q=0.15$, the companion has a mass $2.4M_{\odot}$. Performing disentangling on mock spectra shows that the spectral signature of a non-degenerate stellar companion with such a low mass cannot be retrieved using our data.
   }
   {
   Our measurements do not support the presence of a BH companion in MWC 656. The derived upper limit on the mass of the companion rather indicates that it is a neutron star, a white dwarf, or a hot helium star. Far-UV data will help to reject or confirm a hot helium-star companion.
   }


   \maketitle
% \maketitle

\section{Introduction}
Be stars are suggested to form through binary interactions \citep[e.g.][]{Bodensteiner_et_al_2020_Be,Hastings_et_al_2021,Dallas_et_al_2022, Dufton_et_al_2022}. Therefore, Be-binaries offer unique opportunities to study binary interactions and their products. An important system is MWC\,656, also known as the first reported binary comprising a Be star and a black-hole (BH) companion -- a Be+BH binary \citep[][hereafter referred to as \citetalias{Casares_et_al_2014}]{Casares_et_al_2014}. However, a more recent study refutes the BH scenario \citep[][hereafter referred to as \citetalias{Rivinius_et_al_2022}]{Rivinius_et_al_2022}.

MWC\,656 was suggested to be a $\gamma$-ray binary candidate \citep[][]{Lucarelli_et_al_2010} with an orbital period of 60.37$\pm$0.04\,d derived by photometry \citep{Williams_et_al_2010}. The Be star has an effective temperature $T_{\text{eff}} = 19\,000 \pm 3\,000$\,K \citep{Williams_et_al_2010} and a spectral type B1.5-2\,III (\citetalias{Casares_et_al_2014}). \citet{Williams_et_al_2010} also suggested that the system might be a runaway, given its derived distance of $2.6 \pm 1.0$\,kpc and its high Galactic latitude $b = -12\degree$. 

A spectroscopic study of MWC\,656 revealed the presence of He {\sc ii} $\lambda$4686 in emission (\citetalias{Casares_et_al_2014}), which is suggested to originate from an accretion disc around a hot companion. From an orbital fitting of radial velocity (RV) measurements of the Be star and He {\sc ii} $\lambda$4686, \citetalias{Casares_et_al_2014} derived an eccentricity $e = 0.10\pm0.04$ and a mass ratio $q = M_{\text{comp}}/M_{\mathrm{Be}} = 0.41 \pm 0.07$. With an adopted mass for the Be star between 10\,-\,16\,$M_{\odot}$, they estimate a companion mass of \,4\,-\,7\,$M_{\odot}$. From the absence of spectral features associated with a luminous companion, they concluded that the companion is a BH. No Be+BH binaries were reported in the literature prior to this, as most Be X-ray binaries have confirmed neutron-star (NS) companions or uncertain companions \citep[][]{Belczynski_Ziolkowski_2009}.

MWC 656 was also detected as a faint X-ray source, with luminosities in line with the quiescent Be+BH scenario \citep[][]{Munar-Adrover_et_al_2014, Ribo_et_al_2017}. Furthermore, the combined radio ($L_{\text{8.6GHz}}\sim 10^{26}-10^{27}$\,erg s$^{-1}$) and X-ray luminosities ($L_{\text{1-10keV}}\sim 10^{30}-10^{31}$\,erg s$^{-1}$) are in agreement with those for low-mass quiescent X-ray binaries containing a BH \citep[][]{Dzib_et_al_2015, Ribo_et_al_2017}. As suggested by \citet{Zamanov_et_al_2016}, the BH is expected to constantly accrete from the outer edges of the Be disc, which would give rise to the low detected X-ray luminosities and the quiescent appearance. 

However, a more recent spectroscopic study suggested that MWC 656 hosts a hot subdwarf (sdO) rather than a BH (\citetalias{Rivinius_et_al_2022}). \citetalias{Rivinius_et_al_2022} found a RV semi-amplitude for He\,{\sc ii} $\lambda$4686 similar to \citetalias{Casares_et_al_2014} ($K_{\text{He {\sc ii}}} \sim 80$\,km\,s$^{-1}$). However, the RV semi-amplitudes for the Be star are discrepant.

\citetalias{Casares_et_al_2014} used the Fe\,{\sc ii}\,$\lambda$4583 emission line to obtain RVs for the Be star and found $K_{\text{Be}} = 32.0\pm5.3$\,km\,s$^{-1}$, whereas \citetalias{Rivinius_et_al_2022} used the He\,{\sc i}\,$\lambda$6678 atmospheric absorption line finding $K_{\text{Be}}\,=\,10$ to 15\,km\,s$^{-1}$, thus lowering the mass of the secondary by a factor of about two to three. Hence, \citetalias{Rivinius_et_al_2022} rule out a BH companion. If the companion is indeed not a BH, the number of detected Be+BH binaries falls back to zero, and also the suggestion that Be stars can form through common envelope has to be revised \citep[][]{Grudzinska_et_al_2015}.

In this manuscript, we present a full orbital analysis of MWC 656 based on new high-resolution spectra and we reevaluate the nature of the companion in the system. Section \ref{sec_data} describes the data. Section \ref{sec_analysis} presents the orbital and spectral disentangling analysis. We discuss our results in Sect. \ref{sec_discussion} and conclude in Sect. \ref{sec_conclusions}.

\section{Spectroscopic data}\label{sec_data}
\subsection{New high-resolution data}
New spectroscopic data were obtained with the High-Efficiency and high-Resolution Mercator Echelle Spectrograph (HERMES) instrument, which is mounted on the 1.2-m Mercator Telescope at the Roque de los Muchachos observatory on the Canary island La Palma, Spain \citep[][]{Raskin_2011}. The high-resolution mode has a resolving power of $R = \lambda / \Delta \lambda \simeq 85\,000$ and covers the wavelength range 3800\,-\,9000\,$\AA$ with a step size of $\sim 0.02-0.05\AA$.

We obtained 18 HERMES spectra with signal-to-noise (S/N) ratios around 50-60 per pixel. The dates of observation and S/N of each spectrum can be found in Table \ref{table_spectral_dates}. The normalisation was done with a spline interpolation using the Python package SciPy \citep[][]{SciPy}. Parts of the spectra are shown in Fig. \ref{fig_CaIItriplet}.

\subsection{Archival data}
We also retrieved the 34 archival spectra used by \citetalias{Casares_et_al_2014}. These data were taken with the Fibre-fed RObotic Dual-beam Optical Spectrograph \citep[FRODOspec;][]{Frodospec}, mounted on the 2.0-m Liverpool Telescope (LT), which is also located at the Roque de Los Muchachos Observatory. The data have $R \sim 5500$ and cover the wavelength range 3900\,-\,5215\,$\AA$ with a step size of $\sim 0.4\AA$. Most of the data were taken between April-July 2011 and four spectra were taken in May-June 2012. The data generally have S/N $>200$ per pixel. The normalisation was done similarly to the HERMES spectra.

\begin{table}
    \centering
\caption{Mid-exposure baricentric Julian dates (BJD), exposure times, and S/N of the HERMES spectra of MWC 656. The last column also indicates the corresponding phase in the derived 59.028\,d orbit (see Sect. \ref{sec_orbital_fitting}).}
    \begin{tabular}{ cccc }
     \hline
     \hline
     BJD & exp. time [s] & S/N at H$\alpha$ & phase\\ 
     \hline
      2456227.456 & 900 & 57 & 0.29\\
      2456647.321 & 580 & 53 & 0.40\\
      2457659.462 & 1200 & 56 & 0.55\\
      2457663.402 & 1200 & 49 & 0.61\\
      2459053.691 & 900 & 59 & 0.17\\
      2459057.692 & 900 & 54 & 0.23\\
      2459060.618 & 900 & 59 & 0.28\\
      2459077.606 & 900 & 62 & 0.57\\
      2459081.566 & 900 & 59 & 0.64\\
      2459094.506 & 900 & 62 & 0.86\\
      2459096.467 & 900 & 65 & 0.89\\
      2459102.410 & 900 & 69 & 0.99\\
      2459119.538 & 900 & 56 & 0.28\\
      2459130.500 & 900 & 61 & 0.47\\
      2459384.721 & 600 & 49 & 0.77\\
      2459426.569 & 720 & 62 & 0.48\\
      2459462.614 & 1350 & 51 & 0.09\\
      2459733.687 & 670 & 56 & 0.69\\
     \hline
    \end{tabular}
\label{table_spectral_dates}
\end{table}

\section{Analysis} \label{sec_analysis}
\subsection{Radial velocity measurements} \label{sec_RVs}
We determined the RVs through iterative cross-correlation. By cross-correlating a template over part of the spectra, a cross-correlation function (CCF) is calculated. A parabola fit to the CCF then determines the RV of the spectrum \citep[][]{Zucker_2003}. We also used Gaussian fitting following \citet{Sana_et_al_2013} and obtained very similar results.

For the cross-correlation, the first spectrum was chosen as the initial template. Every iteration creates a master template by coadding the spectra according to the derived RVs. This master template is then used in the next iteration. The obtained RVs are relative to the master template spectrum and hence no systemic velocity is derived. Iterations stopped when RVs converged.

For the Be star, we derived RVs in three different regions: a set of He {\sc i} absorption lines that were fitted simultaneously, the H$\beta$ emission line, and the Fe {\sc ii} $\lambda$4583 emission line (also used by \citetalias{Casares_et_al_2014}). The set of He {\sc i} absorption lines includes He {\sc i} $\lambda\lambda$4026, 4121, 4144, 4388, 4471, 4713, and 5048. We did not include He {\sc i} $\lambda$5876 or He {\sc i} $\lambda$6678 + He {\sc ii} $\lambda$6683 as they seem to have contamination from another emission component (see Fig. \ref{fig_dynamical}, middle and right panel). We also obtained RVs for He {\sc ii} $\lambda$4686. The RVs are listed in Table \ref{table_appendix_RVs} with a 1$\sigma$ error.

\subsection{Orbital fitting}\label{sec_orbital_fitting}
The orbital fitting was performed using the spinOS tool, which uses a non-linear least squares minimisation\footnote{\url{https://github.com/matthiasfabry/spinOS}}. We fitted the orbital period $P$, time of periastron passage $T_0$, eccentricity $e$, argument of periastron $\omega$ (implemented in spinOS with respect to the ascending node), and RV semi-amplitude of the Be star ($K_{\text{Be}}$) and of the companion ($K_{\text{He {\sc ii}}}$). The systemic velocities of each component were also fitted. However, as mentioned in Sect. \ref{sec_RVs}, they have no physical meaning here and we refer to them as the zero-point offsets zo$_{\text{Be}}$ and zo$_{\text{He {\sc ii}}}$. 

The orbital solutions are listed in Table \ref{table_orbital_params} with their 1$\sigma$ error. The first column lists a solution allowing for non-zero eccentricity. However, according to the Lucy-Sweeney test \citep[$e/\sigma_e < 2.4$;][]{Lucy-Sweeney_1971}, the obtained eccentricity is not significant and for the remainder of the manuscript, we assume that the system is circular. 

The circular orbital solutions obtained from the He {\sc i} absorption lines, H$\beta$, and Fe {\sc ii} $\lambda$4583 are all in agreement with each other, suggesting that our results are robust over different spectral lines. Orbital solutions obtained for each of the two components individually are also in agreement (see Table \ref{table_appendix_orbital_params_indiv}). 

We also performed a fit using both the HERMES and LT data. However, when performing cross-correlation on the He {\sc i} absorption lines in the LT data, the RVs show large scatter. While looking at the He {\sc i} absorption lines individually, we noticed that the zero-point of each line is different, probably due to a non-linear wavelength calibration issue in the data. Hence, this forms an issue for fitting multiple lines together. Therefore, we did not include the RVs from the He {\sc i} lines in the LT spectra in the HERMES\,+\,LT orbital fitting. 

\begin{figure}
    \centering
    \includegraphics[width = \linewidth]{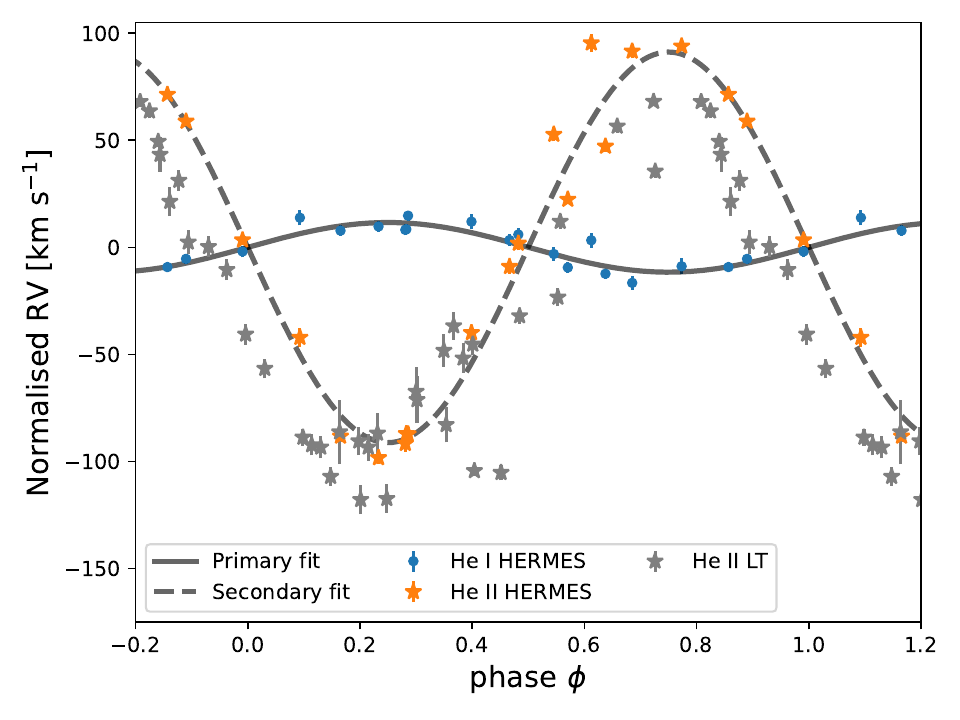}
    \caption{Orbital fit using $P = 59.028$\,d and $T_0$ = 2457509.20. The RVs of each component are subtracted by their respective zo from the HERMES + LT fit.}
    \label{fig_orbital_solution}
\end{figure}

The HERMES\,+\,LT orbital solution is listed in the final column of Table \ref{table_orbital_params} and is shown in Fig. \ref{fig_orbital_solution}. It can be seen that there is an offset between the HERMES and LT He {\sc ii} measurements, potentially attributed due to the wavelength calibration issue in the LT data. Therefore, we fitted an orbital solution including a RV shift $\Delta$RV between the HERMES and LT data using a Levenberg-Marquardt optimisation method \citep[see][]{Sana_et_al_2013}. While fitting for $\Delta$RV, we lower the precision of the semi-amplitudes and mass ratio. However, combining both sets of data does allow for an improvement in $P$ and $T_0$. Therefore, for the remainder of this manuscript, we use only $P$ and $T_0$ from the HERMES\,+\,LT fit and all other parameters from the circular fit on the He {\sc i} lines, since the use of absorption lines is generally preferred over emission lines.

The orbital period that we derived ($P \simeq 59$\,d) is slightly but significantly smaller than the one used by \citetalias{Casares_et_al_2014}. The derived mass ratio of $q = 0.12 \pm 0.03$, is much lower than the previously reported value of $q = 0.41\pm0.07$ (\citetalias{Casares_et_al_2014}). Combining our mass ratio with a mass of $7.8\pm 2.0 M_{\odot}$ for the Be star \citep[][]{Williams_et_al_2010}, the mass of the companion is $0.94 \pm 0.34 M_{\odot}$. This mass range is in agreement with a low-mass main-sequence star, white dwarf (WD), and sdO companion. However, it is not massive enough for a BH ($M \gtrsim 2.5 M_{\odot}$). Even when the mass of the Be star would be 16\,$M_{\odot}$, which is the upper limit proposed by \citetalias{Casares_et_al_2014}, and the mass ratio would be 0.15, the mass of the companion would be 2.4$M_{\odot}$ and hence still not in the BH-mass range. Therefore, it is highly unlikely that MWC 656 hosts a BH. However, a neutron-star (NS) companion seems possible. We further discuss the different possibilities in Sect. \ref{sec_nature_companion}. 

Our derived orbital solutions for He {\sc i} and Fe {\sc ii} agree with each other and with the RV semi-amplitude mentioned in \citetalias{Rivinius_et_al_2022}. To investigate the cause of the discrepancy between the RV semi-amplitude of Fe {\sc ii} ($K_{\text{Fe {\sc ii}}}$) derived here and the one derived by \citetalias{Casares_et_al_2014}, we also measured RVs for Fe {\sc ii} $\lambda$4583 in the LT data using both cross-correlation and double-Gaussian fitting. In both cases, we find $K_{\text{Fe {\sc ii}}} \approx 18\,km\,s^{-1}$, with a larger dispersion (3\,km\,s$^{-1}$ versus 4\,km\,s$^{-1}$) for the Gaussian-fitting method. This is in contrast with both the value listed in \citetalias{Casares_et_al_2014} and the one obtained from the HERMES data. The discrepancy with the latter might be due to the resolution of the data, however, HERMES and LT agree within 1.5$\sigma$. We currently cannot explain the discrepancy with \citetalias{Casares_et_al_2014}.

\begin{table*}
    \centering
\caption{Orbital and dynamical parameters of MWC 656. Only the last column includes LT data in the fit.}
    \begin{tabular}{ ccccc|c }
     \hline
     \hline
     used Be line: & \multicolumn{2}{c}{He {\sc i} absorption lines$^{*}$} & H$\beta$& Fe {\sc ii} $\lambda$4583 & HERMES + LT (He {\sc ii}) \\ 
     \hline
      $P$ [d] & 59.085 $\pm$ 0.026 & 59.094 $\pm$ 0.024 & 59.090 $\pm$ 0.033 & 59.094 $\pm$ 0.029 & 59.028 $\pm$ 0.011\\
      $T_0$ [bjd-2450000] & 7510.0 $\pm$ 8.5 & 7507.53 $\pm$ 0.61 & 7507.69 $\pm$ 0.84 & 7507.58 $\pm$ 0.75 & 7509.20 $\pm$ 0.31\\
      e & 0.037 $\pm$ 0.035 & 0 (fixed) & 0 (fixed) & 0 (fixed) & 0 (fixed)\\
      $\omega$ [$\degree$] & 105 $\pm$ 51 & 90 (fixed) & 90 (fixed) & 90 (fixed) & 90 (fixed)\\
      $K_{\mathrm{Be}}$ [km s$^{-1}$] & 11.5 $\pm$ 2.8 & 11.5 $\pm$ 2.8 & 9.8 $\pm$ 2.0 & 11.7 $\pm$ 2.6 & 11.6 $\pm$ 3.7\\
      $K_{\text{He {\sc ii}}}$ [km s$^{-1}$] & 92.6 $\pm$ 3.1 & 92.9 $\pm$ 3.0 & 92.7 $\pm$ 4.2 & 92.8 $\pm$ 3.7 & 91.2 $\pm$ 2.9\\
      $q = K_1/K_{\text{He {\sc ii}}}$ & 0.12 $\pm$ 0.03 & 0.12 $\pm$ 0.03 & 0.11 $\pm$ 0.02 & 0.13 $\pm$ 0.03 & 0.13 $\pm$ 0.05\\
      zo$_{\text{Be}}$ [km s$^{-1}$] & $-$14.5 $\pm$ 2.1 & $-$14.5 $\pm$ 2.1 & $-$13.7 $\pm$ 1.4 & $-$32.0 $\pm$ 1.9 & $-$14.8 $\pm$ 2.8\\
      zo$_{\text{He {\sc ii}}}$ [km s$^{-1}$] & 85.1 $\pm$ 2.3 & 85.2 $\pm$ 2.2 & 85.3 $\pm$ 3.1 & 85.2 $\pm$ 2.7 & 87.2 $\pm$ 2.8\\
      $\Delta$RV [km s$^{-1}$] & /&/&/&/& 30.0 $\pm$ 4.1\\
      reduced $\chi^2$ & 10.6 & 10.3 & 19.8 & 15.5 & 18.7\\
     \hline
    \end{tabular}
         \flushleft
    \begin{tablenotes}
      \small
      \item * lines included in the cross-correlation range: He {\sc i} $\lambda\lambda$4026, 4121, 4144, 4388, 4471, 4713, 5048
    \end{tablenotes}
\label{table_orbital_params}
\end{table*}
\begin{table}
    \centering
\caption{Orbital and dynamical parameters obtained from each component individually.}
    \begin{tabular}{ ccc }
     \hline
     \hline
     used line: & He {\sc i} absorption$^{*}$ & He {\sc ii} $\lambda$4686 \\ 
     \hline
      $P$ [d] & 58.906 $\pm$ 0.092 & 59.098 $\pm$ 0.032\\
      $T_0$ [bjd-2450000] & 7511.7 $\pm$ 2.4 & 7477.89 $\pm$ 0.84$^{**}$\\
      e & 0 (fixed) &  0 (fixed)\\
      $\omega$ [$\degree$] & 90 (fixed) & 90 (fixed)\\
      $K$ [km s$^{-1}$] & 11.8 $\pm$ 1.3 & 93.0 $\pm$ 4.1 \\
      % $K$ [km s$^{-1}$] & / & 92.954 $\pm$ 4.09\\
      zo [km s$^{-1}$] & $-15.08 \pm 0.94$ & $85.0\pm3.0$\\
      reduced $\chi^2$ & 3.9 & 18.9\\
     \hline
    \end{tabular}
         \flushleft
    \begin{tablenotes}
      \small
      \item * lines included in the cross-correlation range: He {\sc i} $\lambda\lambda$4026, 4121, 4144, 4388, 4471, 4713, 5048\\
      \item ** The $\sim$30 day difference with respect to all other orbital solutions originates in using the secondary as the primary 
    \end{tablenotes}
\label{table_appendix_orbital_params_indiv}
\end{table}

\subsection{Spectral disentangling}\label{sec_disentangling_results}
Spectral disentangling can reveal the signature of companions two orders of magnitudes fainter than the primary star with a good quality data set. Thus, we can push the detection of companions down to very low light ratios \citep[see e.g.][]{Bodensteiner_et_al_2020, Shenar_et_al_2020, Shenar_et_al_2022}.

\begin{figure*}
    \centering
    \includegraphics[width = \linewidth]{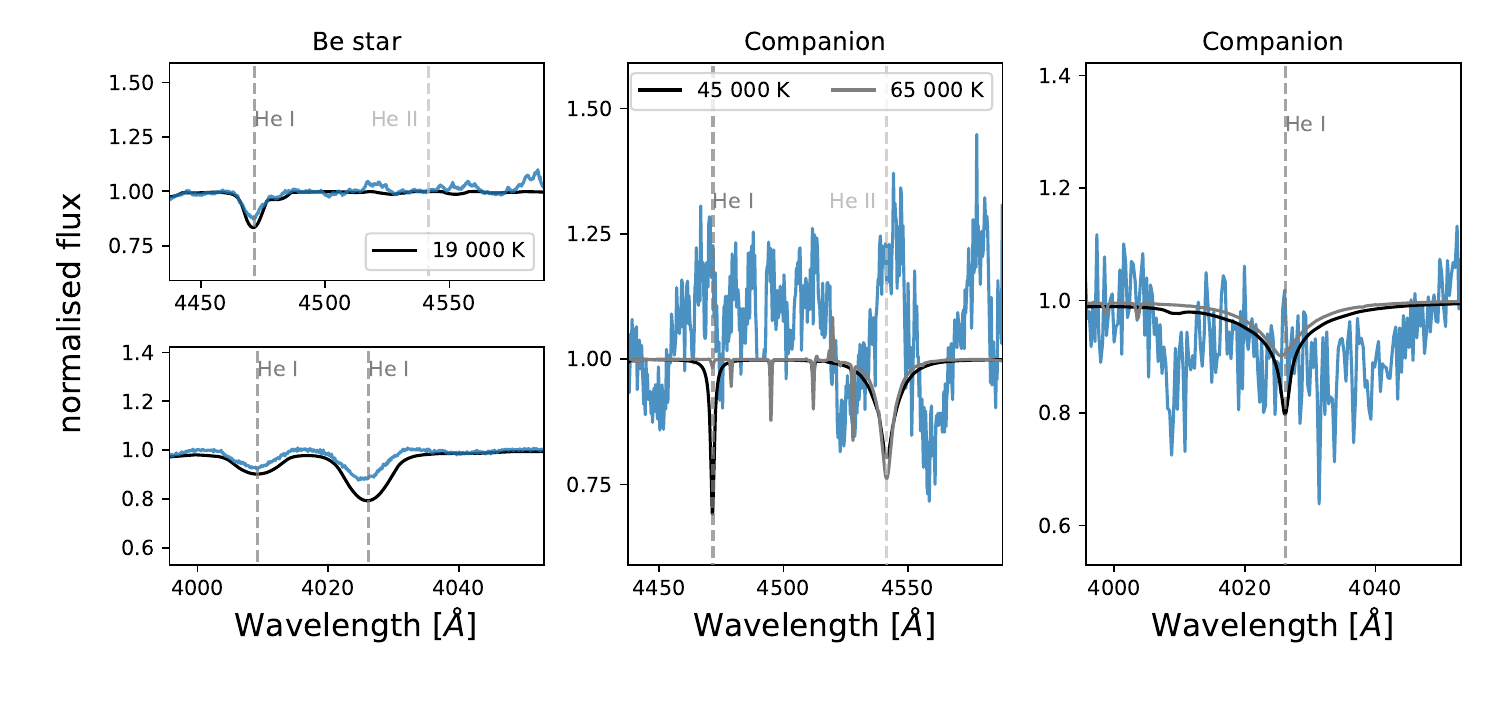}
    \caption{Disentangled spectrum (blue) for the primary (Be star, left panels) and the companion (middle and right panel). The model in the left panels (black) is for a B star of 19\,000\,K with $v\sin i = 330$\,km s$^{-1}$. The models in the middle and right panels are for a stripped star of 45\,000\,K (black) and 65\,000\,K (grey) with $v\sin i = 30$\,km s$^{-1}$. Emission features in the top-left panel are mostly due to iron.}
    \label{fig_disentangled_data}
\end{figure*}

We used the iterative shift-and-add technique \citep[][]{Gonzalez-Levato_2006, Shenar_et_al_2020, Shenar_et_al_2022}\footnote{\url{https://github.com/TomerShenar/Disentangling_Shift_And_Add}}, fixing the orbital parameters to the values we derived in Table \ref{table_orbital_params}. Figure \ref{fig_disentangled_data} shows the disentangled spectra zoomed in on the regions of interest. 

Due to strong variability in the emission lines, we avoid regions such as the Balmer and Paschen lines, and the He {\sc ii} $\lambda$4686 line. While excluding Balmer lines from the disentangling, we cannot detect a WD companion. Even if we would be able to disentangle the Balmer lines, a WD would contribute an estimated $\sim 0.04\%$ to the optical spectra, which is too faint to be detected with the current data.

To determine whether a stripped star could be detected in this system, we performed disentangling on mock spectra. However, these simulations showed that neither a lowest-mass (1\,$M_{\odot}$) nor a most massive ($\sim$ 2.5\,$M_{\odot}$) stripped star can be unambiguously detected (see Appendix \ref{appendix_spectral_disentangling}). Indeed, no clear signature of a companion is detected in our data (Fig. \ref{fig_disentangled_data}).

\section{Discussion}\label{sec_discussion}

\subsection{Does the He\,{\sc ii} emission line originate from a disc around the companion?}
The shape of H$\alpha$ is very variable, often differing from a double peak, which is indicative of a perturbed outer disc \citep[see e.g.][]{Zamanov_et_al_2022}. This supports the estimation that the companion is orbiting through the outer edges of the Be-disk \citep[][]{Zamanov_et_al_2016}, implying that an accretion disk may form.

In principle, it could also be the case that the hot (see Sect. \ref{sec_nature_companion}) companion orbiting around the Be star heats up part of the Be disc \citep[see also fig. 8 in][for the example of FY CMa]{Peters_et_al_2008}. In this scenario, the He\,{\sc ii} $\lambda$4686 (and He\,{\sc i} $\lambda$6678 + He {\sc ii} $\lambda$6683 and He\,{\sc i} $\lambda$5875) emission line form in the Be disc. Hence, the obtained RVs from the He\,{\sc ii} line could not be used to constrain the mass of the companion.

While He\,{\sc ii} is variable in shape, several epochs show a double-peaked feature (e.g. the grey spectrum in the top-right panel of Fig. \ref{fig_CaIItriplet}), suggesting a Keplerian disc. If part of the Be disc would be heated up, it seems rather unlikely that the shape of the emission line resembles that of a Keplerian disc.

For the same reason, the He emission lines most likely not originate from stellar winds. However, if it were stellar winds, the derived RVs would still trace the motion of the hot companion. Therefore, we suggest that the He emission lines originate from an accretion disc around the hot companion.

\subsection{The nature of the unseen companion}\label{sec_nature_companion}
In Sect. \ref{sec_orbital_fitting}, we found that the companion is either a low-mass main-sequence star, WD, sdO, or NS. The presence of He {\sc ii} emission indicates a hot companion. A low-mass main-sequence companion with a mass $< 2.5 M_{\odot}$ would have $T_{\text{eff}} \lesssim 10\,000$\,K and is thus far too cool to be responsible for the He {\sc ii} emission. Therefore, we can exclude this possibility.

Any viable scenario must also be able to explain the system's radio and X-ray emission \citep[][]{Munar-Adrover_et_al_2014,Dzib_et_al_2015,Ribo_et_al_2017}. Be stars can exhibit X-ray luminosities of $\sim 10^{29}-10^{31}$\,erg s$^{-1}$ \citep[e.g.,][]{Naze_et_al_2018} and radio luminosities of $\sim 10^{26}-10^{27}$\,erg s$^{-1}$ \citep[e.g.,][]{Taylor1990} that can originate from the Be disc or wind. The presence of an sdO or accreting compact companion can also cause X-ray and radio emission in the range observed for MWC 656.

Below we investigate whether a WD, sdO, or NS companion agree with the observed spectral features as well as the detected X-ray and radio. We do not discuss the $\gamma$-rays in detail, since it might be attributed to another source \citep[see][]{Alexander_et_al_2015,Munar-Adrover_et_al_2016}.

\subsubsection{A white dwarf}
If the companion would be a WD, it could be a candidate for a type Ia supernova progenitor. While the companion is currently accreting at low levels from the disc, the accretion rate might significantly increase during the evolution of the Be star. From an evolutionary perspective, the WD scenario seems the most likely, with an estimated 70\% of all Be stars having a WD companion \citep[][]{van_Bever_Vanbeveren_1997,Raguzova_2001}. While no Be\,+\,WD binaries are found yet in the Milky Way, some are reported in the Magellanic Clouds \citep[see][]{Kahabka_et_al_2006, Sturm_et_al_2012, Coe_et_al_2020}. Furthermore, the presence of He\,{\sc ii} $\lambda$4686 emission was already reported in several binaries containing accreting WDs \citep[such as in (magnetic) cataclysmic variables, e.g.][]{Oliviera_et_al_2017, Mason_et_al_2019, Hou_et_al_2023}. However, it seems that these systems show Balmer emission at least as strong as the He {\sc ii} $\lambda$4686 emission, which does not seem to be the case here. The X-ray and radio luminosity of MWC 656 are both in the realm of those detected for the brightest accreting (magnetic) white dwarfs \citep[][]{Coppejans_Knigge_2020,Hewitt_et_al_2020}.

\subsubsection{A hot subdwarf}
MWC 656 shows striking similarities with previously detected Be\,+\,sdO binaries (such as e.g. $\phi$ Per \citeauthor{Poeckert_1981} \citeyear{Poeckert_1981}, FY CMa \citeauthor{Peters_et_al_2008} \citeyear{Peters_et_al_2008}, $o$ Pup \citeauthor{Koubski_et_al_2012} \citeyear{Koubski_et_al_2012}, HD 55606 \citeauthor{Chojnowski_et_al_2018} \citeyear{Chojnowski_et_al_2018}). MWC 656 shows strong Ca {\sc ii} $\lambda\lambda$8202, 8249, 8255 emission lines (see bottom panel of Fig. \ref{fig_CaIItriplet}), also detected in $o$ Pup and HD 55606. Furthermore, just like in several studied Be\,+\,sdO binaries, there is additional emission visible in He\,{\sc i} $\lambda$6678 + He {\sc ii} $\lambda$ 6683 and He\,{\sc i} $\lambda$ 5875 (Fig. \ref{fig_dynamical} top, middle and right panels) which trace the same RV motion as the He {\sc ii} $\lambda$4686 emission line (Fig. \ref{fig_dynamical} top-left panel). So far, He {\sc ii} $\lambda$4686 emission was only detected in $\phi$ Per and 59 Cyg \citep[][]{Rivinius_Stefl_2000}. In both cases, the emission is very weak.

The X-ray luminosity of MWC 656 is in line with that observed for known Be\,+\,sdO binaries \citep[][]{Naze_et_al_2020}. The detected radio luminosity also roughly agrees with other Be\,+\,sdO systems \citep[][]{Wendker_1995,Wendker_2015}.

\subsubsection{A neutron star companion}
As suggested by \citet{Williams_et_al_2010}, MWC\,656 might be a runaway. This could indicate a past supernova kick from a companion. However, a kick is also expected to introduce eccentricity. The circular orbit could be explained by an electron-capture supernova \citep[][]{Podsiadlowksi_et_al_2004}. In this scenario, the potential runaway status of MWC 656 is probably due to a dynamical ejection.

The X-ray spectrum of MWC 656 \citep[][]{Munar-Adrover_et_al_2014}, consisting of a soft thermal emission component ($kT_{bb}\sim$0.1~keV) and a harder emission tail (described as a $\Gamma \sim 1$ powerlaw), is 
not unlike that of neutron stars Be X-ray binaries observed at similarly low X-ray luminosities \citep[e.g.][]{Rouco-Escorial_et_al_2019,Tsyganov_et_al_2020}. The radio emission from Be\,+\,NS systems has not been explored in this regime \citep[][]{van-den-eijnden_et_al_2021}.

\section{Conclusion}\label{sec_conclusions}
Based on new high-resolution spectroscopic data, we investigated the nature of the companion to the Be star in the system MWC 656. We find a mass ratio of $q = 0.12 \pm 0.03$, much lower than the previously reported $q = 0.41 \pm 0.07$.

A BH companion is thus ruled out by our analysis, in line with the findings of \citetalias{Rivinius_et_al_2022}. We suggest that this discrepancy with \citetalias{Casares_et_al_2014} can be due to both the higher resolution of the data as well as the method used to derive the RVs. 

The most likely companions are a sdO, WD, or NS. On the one hand, the circular orbit might indicate a sdO or WD companion. While on the other hand the position on the sky might indicate a runaway and hence favor the NS scenario. Given the spectroscopic, X-ray, and radio data, we do not favor any possibility more than the other.

All three companions would be equally exciting. If MWC 656 hosts
\begin{itemize}
    \item a NS, the low eccentricity is very unique and hints towards an electron-capture supernova,
    \item a sdO, it adds one more to the very few known Be+sdO binaries, which are extremely valuable to study close binary interactions,
    \item a WD, it might be a Type Ia supernova candidate.
\end{itemize}
Upcoming far-UV data could tell us more about the nature of the companion, giving the possibility to either confirm or reject at least an sdO companion. 
% \newpage~\newpage

\begin{acknowledgements}
The authors thank Jaime Villaseñor and Jan Van Roestel for the insightful and interesting discussions regarding this system. We also thank Joey Mombarg and Sergio Simón-Díaz for contributing to the observational campaign.

SJ acknowledges support from the FWO PhD fellowship under project 11E1721N. 

This project has received funding from the European Research Council (ERC) under the European Union's Horizon 2020 research and innovation programme (grant agreement n$\degree$ 772225/MULTIPLES). TS acknowledges support from the European Union's Horizon 2020 under the Marie Skłodowska-Curie grant agreement No 101024605. The research leading to these results has received funding from the European Research Council (ERC) under the European Union's Horizon 2020 research and innovation programme (grant agreement N$^\circ$670519: MAMSIE) and from the KU Leuven Research Council (grant C16/18/005: PARADISE).
\end{acknowledgements}

% \printbibliography
\bibliography{refs}

\begin{thebibliography}{54}
\expandafter\ifx\csname natexlab\endcsname\relax\def\natexlab#1{#1}\fi

\bibitem[{{Alexander} \& {McSwain}(2015)}]{Alexander_et_al_2015}
{Alexander}, M.~J. \& {McSwain}, M.~V. 2015, \mnras, 449, 1686

\bibitem[{{Belczynski} \& {Ziolkowski}(2009)}]{Belczynski_Ziolkowski_2009}
{Belczynski}, K. \& {Ziolkowski}, J. 2009, \apj, 707, 870

\bibitem[{{Bodensteiner} {et~al.}(2020{\natexlab{a}}){Bodensteiner}, {Shenar},
  {Mahy}, {Fabry}, {Marchant}, {Abdul-Masih}, {Banyard}, {Bowman}, {Dsilva},
  {Frost}, {Hawcroft}, {Reggiani}, \& {Sana}}]{Bodensteiner_et_al_2020}
{Bodensteiner}, J., {Shenar}, T., {Mahy}, L., {et~al.} 2020{\natexlab{a}},
  \aap, 641, A43

\bibitem[{{Bodensteiner} {et~al.}(2020{\natexlab{b}}){Bodensteiner}, {Shenar},
  \& {Sana}}]{Bodensteiner_et_al_2020_Be}
{Bodensteiner}, J., {Shenar}, T., \& {Sana}, H. 2020{\natexlab{b}}, \aap, 641,
  A42

\bibitem[{{Casares} {et~al.}(2014){Casares}, {Negueruela}, {Rib{\'o}}, {Ribas},
  {Paredes}, {Herrero}, \& {Sim{\'o}n-D{\'\i}az}}]{Casares_et_al_2014}
{Casares}, J., {Negueruela}, I., {Rib{\'o}}, M., {et~al.} 2014, \nat, 505, 378

\bibitem[{{Chojnowski} {et~al.}(2018){Chojnowski}, {Labadie-Bartz}, {Rivinius},
  {Gies}, {Panoglou}, {Borges Fernandes}, {Wisniewski}, {Whelan}, {Mennickent},
  {McMillan}, {Dembicky}, {Gray}, {Rudyk}, {Stringfellow}, {Lester},
  {Hasselquist}, {Zharikov}, {Levenhagen}, {Souza}, {Leister}, {Stassun},
  {Siverd}, \& {Majewski}}]{Chojnowski_et_al_2018}
{Chojnowski}, S.~D., {Labadie-Bartz}, J., {Rivinius}, T., {et~al.} 2018, \apj,
  865, 76

\bibitem[{{Coe} {et~al.}(2020){Coe}, {Kennea}, {Evans}, \&
  {Udalski}}]{Coe_et_al_2020}
{Coe}, M.~J., {Kennea}, J.~A., {Evans}, P.~A., \& {Udalski}, A. 2020, \mnras,
  497, L50

\bibitem[{{Coppejans} \& {Knigge}(2020)}]{Coppejans_Knigge_2020}
{Coppejans}, D.~L. \& {Knigge}, C. 2020, \nar, 89, 101540

\bibitem[{{Dallas} {et~al.}(2022){Dallas}, {Oey}, \&
  {Castro}}]{Dallas_et_al_2022}
{Dallas}, M.~M., {Oey}, M.~S., \& {Castro}, N. 2022, \apj, 936, 112

\bibitem[{{Dufton} {et~al.}(2022){Dufton}, {Lennon}, {Villase{\~n}or},
  {Howarth}, {Evans}, {de Mink}, {Sana}, \& {Taylor}}]{Dufton_et_al_2022}
{Dufton}, P.~L., {Lennon}, D.~J., {Villase{\~n}or}, J.~I., {et~al.} 2022,
  \mnras, 512, 3331

\bibitem[{{Dzib} {et~al.}(2015){Dzib}, {Massi}, \& {Jaron}}]{Dzib_et_al_2015}
{Dzib}, S.~A., {Massi}, M., \& {Jaron}, F. 2015, \aap, 580, L6

\bibitem[{{Gonz{\'a}lez} \& {Levato}(2006)}]{Gonzalez-Levato_2006}
{Gonz{\'a}lez}, J.~F. \& {Levato}, H. 2006, \aap, 448, 283

\bibitem[{{G{\"o}tberg} {et~al.}(2018){G{\"o}tberg}, {de Mink}, {Groh},
  {Kupfer}, {Crowther}, {Zapartas}, \& {Renzo}}]{Gotberg_et_al_2018}
{G{\"o}tberg}, Y., {de Mink}, S.~E., {Groh}, J.~H., {et~al.} 2018, \aap, 615,
  A78

\bibitem[{{Grudzinska} {et~al.}(2015){Grudzinska}, {Belczynski}, {Casares}, {de
  Mink}, {Ziolkowski}, {Negueruela}, {Rib{\'o}}, {Ribas}, {Paredes}, {Herrero},
  \& {Benacquista}}]{Grudzinska_et_al_2015}
{Grudzinska}, M., {Belczynski}, K., {Casares}, J., {et~al.} 2015, \mnras, 452,
  2773

\bibitem[{{Hamann} \& {Gr{\"a}fener}(2003)}]{PoWR}
{Hamann}, W.~R. \& {Gr{\"a}fener}, G. 2003, \aap, 410, 993

\bibitem[{{Hastings} {et~al.}(2021){Hastings}, {Langer}, {Wang},
  {Schootemeijer}, \& {Milone}}]{Hastings_et_al_2021}
{Hastings}, B., {Langer}, N., {Wang}, C., {Schootemeijer}, A., \& {Milone},
  A.~P. 2021, \aap, 653, A144

\bibitem[{{Hewitt} {et~al.}(2020){Hewitt}, {Pretorius}, {Woudt}, {Tremou},
  {Miller-Jones}, {Knigge}, {Castro Segura}, {Williams}, {Fender}, {Armstrong},
  {Groot}, {Heywood}, {Horesh}, {van der Horst}, {Koerding}, {McBride},
  {Mooley}, {Rowlinson}, {Stappers}, \& {Wijers}}]{Hewitt_et_al_2020}
{Hewitt}, D.~M., {Pretorius}, M.~L., {Woudt}, P.~A., {et~al.} 2020, \mnras,
  496, 2542

\bibitem[{{Hou} {et~al.}(2023){Hou}, {Luo}, {Dong}, {Chen}, \&
  {Bai}}]{Hou_et_al_2023}
{Hou}, W., {Luo}, A.~L., {Dong}, Y.-Q., {Chen}, X.-L., \& {Bai}, Z.-R. 2023,
  \aj, 165, 148

\bibitem[{{Kahabka} {et~al.}(2006){Kahabka}, {Haberl}, {Payne}, \&
  {Filipovi{\'c}}}]{Kahabka_et_al_2006}
{Kahabka}, P., {Haberl}, F., {Payne}, J.~L., \& {Filipovi{\'c}}, M.~D. 2006,
  \aap, 458, 285

\bibitem[{{Koubsk{\'y}} {et~al.}(2012){Koubsk{\'y}}, {Kotkov{\'a}}, {Votruba},
  {{\v{S}}lechta}, \& {Dvo{\v{r}}{\'a}kov{\'a}}}]{Koubski_et_al_2012}
{Koubsk{\'y}}, P., {Kotkov{\'a}}, L., {Votruba}, V., {{\v{S}}lechta}, M., \&
  {Dvo{\v{r}}{\'a}kov{\'a}}, {\v{S}}. 2012, \aap, 545, A121

\bibitem[{{Lanz} \& {Hubeny}(2007)}]{TLUSTY}
{Lanz}, T. \& {Hubeny}, I. 2007, \apjs, 169, 83

\bibitem[{{Lucarelli} {et~al.}(2010){Lucarelli}, {Verrecchia}, {Striani},
  {Pittori}, {Tavani}, {Vercellone}, {Bulgarelli}, {Gianotti}, {Trifoglio},
  {Chen}, {Giuliani}, {Mereghetti}, {Caraveo}, {Perotti}, {Donnarumma},
  {D'Ammando}, {Del Monte}, {Evangelista}, {Feroci}, {Lazzarotto}, {Pacciani},
  {Soffitta}, {Costa}, {Lapshov}, {Rapisarda}, {Argan}, {Piano}, {Pucella},
  {Sabatini}, {Trois}, {Vittorini}, {Fuschino}, {Galli}, {Labanti},
  {Marisaldi}, {Di Cocco}, {Pellizzoni}, {Pilia}, {Barbiellini}, {Longo},
  {Moretti}, {Vallazza}, {Morselli}, {Picozza}, {Prest}, {Lipari}, {Zanello},
  {Cattaneo}, {Rappoldi}, {Santolamazza}, {Colafrancesco}, {Giommi}, \&
  {Salotti}}]{Lucarelli_et_al_2010}
{Lucarelli}, F., {Verrecchia}, F., {Striani}, E., {et~al.} 2010, The
  Astronomer's Telegram, 2761, 1

\bibitem[{{Lucy} \& {Sweeney}(1971)}]{Lucy-Sweeney_1971}
{Lucy}, L.~B. \& {Sweeney}, M.~A. 1971, \aj, 76, 544

\bibitem[{{Mason} {et~al.}(2019){Mason}, {Wells}, {Motsoaledi}, {Szkody}, \&
  {Gonzalez}}]{Mason_et_al_2019}
{Mason}, P.~A., {Wells}, N.~K., {Motsoaledi}, M., {Szkody}, P., \& {Gonzalez},
  E. 2019, \mnras, 488, 2881

\bibitem[{{Morales-Rueda} {et~al.}(2004){Morales-Rueda}, {Carter}, {Steele},
  {Charles}, \& {Worswick}}]{Frodospec}
{Morales-Rueda}, L., {Carter}, D., {Steele}, I.~A., {Charles}, P.~A., \&
  {Worswick}, S. 2004, Astronomische Nachrichten, 325, 215

\bibitem[{{Munar-Adrover} {et~al.}(2014){Munar-Adrover}, {Paredes}, {Rib{\'o}},
  {Iwasawa}, {Zabalza}, \& {Casares}}]{Munar-Adrover_et_al_2014}
{Munar-Adrover}, P., {Paredes}, J.~M., {Rib{\'o}}, M., {et~al.} 2014, \apjl,
  786, L11

\bibitem[{{Munar-Adrover} {et~al.}(2016){Munar-Adrover}, {Sabatini}, {Piano},
  {Tavani}, {Nguyen}, {Lucarelli}, {Verrecchia}, \&
  {Pittori}}]{Munar-Adrover_et_al_2016}
{Munar-Adrover}, P., {Sabatini}, S., {Piano}, G., {et~al.} 2016, \apj, 829, 101

\bibitem[{{Naz{\'e}} \& {Motch}(2018)}]{Naze_et_al_2018}
{Naz{\'e}}, Y. \& {Motch}, C. 2018, \aap, 619, A148

\bibitem[{{Naz{\'e}} {et~al.}(2022){Naz{\'e}}, {Rauw}, {Smith}, \&
  {Motch}}]{Naze_et_al_2020}
{Naz{\'e}}, Y., {Rauw}, G., {Smith}, M.~A., \& {Motch}, C. 2022, \mnras, 516,
  3366

\bibitem[{{Oliveira} {et~al.}(2017){Oliveira}, {Rodrigues}, {Cieslinski},
  {Jablonski}, {Silva}, {Almeida}, {Rodr{\'\i}guez-Ardila}, \&
  {Palhares}}]{Oliviera_et_al_2017}
{Oliveira}, A.~S., {Rodrigues}, C.~V., {Cieslinski}, D., {et~al.} 2017, \aj,
  153, 144

\bibitem[{{Peters} {et~al.}(2008){Peters}, {Gies}, {Grundstrom}, \&
  {McSwain}}]{Peters_et_al_2008}
{Peters}, G.~J., {Gies}, D.~R., {Grundstrom}, E.~D., \& {McSwain}, M.~V. 2008,
  \apj, 686, 1280

\bibitem[{{Podsiadlowski} {et~al.}(2004){Podsiadlowski}, {Langer},
  {Poelarends}, {Rappaport}, {Heger}, \& {Pfahl}}]{Podsiadlowksi_et_al_2004}
{Podsiadlowski}, P., {Langer}, N., {Poelarends}, A.~J.~T., {et~al.} 2004, \apj,
  612, 1044

\bibitem[{{Poeckert}(1981)}]{Poeckert_1981}
{Poeckert}, R. 1981, \pasp, 93, 297

\bibitem[{{Raguzova}(2001)}]{Raguzova_2001}
{Raguzova}, N.~V. 2001, \aap, 367, 848

\bibitem[{{Raskin} {et~al.}(2011){Raskin}, {van Winckel}, {Hensberge},
  {Jorissen}, {Lehmann}, {Waelkens}, {Avila}, {de Cuyper}, {Degroote},
  {Dubosson}, {Dumortier}, {Fr{\'e}mat}, {Laux}, {Michaud}, {Morren}, {Perez
  Padilla}, {Pessemier}, {Prins}, {Smolders}, {van Eck}, \&
  {Winkler}}]{Raskin_2011}
{Raskin}, G., {van Winckel}, H., {Hensberge}, H., {et~al.} 2011, \aap, 526, A69

\bibitem[{{Rib{\'o}} {et~al.}(2017){Rib{\'o}}, {Munar-Adrover}, {Paredes},
  {Marcote}, {Iwasawa}, {Mold{\'o}n}, {Casares}, {Migliari}, \&
  {Paredes-Fortuny}}]{Ribo_et_al_2017}
{Rib{\'o}}, M., {Munar-Adrover}, P., {Paredes}, J.~M., {et~al.} 2017, \apjl,
  835, L33

\bibitem[{{Rivinius} {et~al.}(2022){Rivinius}, {Klement}, {Chojnowski},
  {Baade}, {Shepard}, \& {Hadrava}}]{Rivinius_et_al_2022}
{Rivinius}, T., {Klement}, R., {Chojnowski}, S.~D., {et~al.} 2022, arXiv
  e-prints, arXiv:2208.12315

\bibitem[{{Rivinius} \& {{\v{S}}tefl}(2000)}]{Rivinius_Stefl_2000}
{Rivinius}, T. \& {{\v{S}}tefl}, S. 2000, in Astronomical Society of the
  Pacific Conference Series, Vol. 214, IAU Colloq. 175: The Be Phenomenon in
  Early-Type Stars, ed. M.~A. {Smith}, H.~F. {Henrichs}, \& J.~{Fabregat}, 581

\bibitem[{{Rouco Escorial} {et~al.}(2019){Rouco Escorial}, {Wijnands}, {Ootes},
  {Degenaar}, {Snelders}, {Kaper}, {Cackett}, \&
  {Homan}}]{Rouco-Escorial_et_al_2019}
{Rouco Escorial}, A., {Wijnands}, R., {Ootes}, L.~S., {et~al.} 2019, \aap, 630,
  A105

\bibitem[{{Sana} {et~al.}(2013){Sana}, {Le Bouquin}, {Mahy}, {Absil}, {De
  Becker}, \& {Gosset}}]{Sana_et_al_2013}
{Sana}, H., {Le Bouquin}, J.~B., {Mahy}, L., {et~al.} 2013, \aap, 553, A131

\bibitem[{{Shenar} {et~al.}(2020){Shenar}, {Bodensteiner}, {Abdul-Masih},
  {Fabry}, {Mahy}, {Marchant}, {Banyard}, {Bowman}, {Dsilva}, {Hawcroft},
  {Reggiani}, \& {Sana}}]{Shenar_et_al_2020}
{Shenar}, T., {Bodensteiner}, J., {Abdul-Masih}, M., {et~al.} 2020, \aap, 639,
  L6

\bibitem[{{Shenar} {et~al.}(2022){Shenar}, {Sana}, {Mahy}, {Ma{\'\i}z
  Apell{\'a}niz}, {Crowther}, {Gromadzki}, {Herrero}, {Langer}, {Marchant},
  {Schneider}, {Sen}, {Soszy{\'n}ski}, \& {Toonen}}]{Shenar_et_al_2022}
{Shenar}, T., {Sana}, H., {Mahy}, L., {et~al.} 2022, \aap, 665, A148

\bibitem[{{Sturm} {et~al.}(2012){Sturm}, {Haberl}, {Pietsch}, {Coe},
  {Mereghetti}, {La Palombara}, {Owen}, \& {Udalski}}]{Sturm_et_al_2012}
{Sturm}, R., {Haberl}, F., {Pietsch}, W., {et~al.} 2012, \aap, 537, A76

\bibitem[{{Taylor} {et~al.}(1990){Taylor}, {Waters}, {Bjorkman}, \&
  {Dougherty}}]{Taylor1990}
{Taylor}, A.~R., {Waters}, L.~B.~F.~M., {Bjorkman}, K.~S., \& {Dougherty},
  S.~M. 1990, \aap, 231, 453

\bibitem[{{Tsygankov} {et~al.}(2020){Tsygankov}, {Doroshenko}, {Mushtukov},
  {Haberl}, {Vasilopoulos}, {Maitra}, {Santangelo}, {Lutovinov}, \&
  {Poutanen}}]{Tsyganov_et_al_2020}
{Tsygankov}, S.~S., {Doroshenko}, V., {Mushtukov}, A.~A., {et~al.} 2020, \aap,
  637, A33

\bibitem[{{van Bever} \& {Vanbeveren}(1997)}]{van_Bever_Vanbeveren_1997}
{van Bever}, J. \& {Vanbeveren}, D. 1997, \aap, 322, 116

\bibitem[{{van den Eijnden} {et~al.}(2021){van den Eijnden}, {Degenaar},
  {Russell}, {Wijnands}, {Bahramian}, {Miller-Jones}, {Hern{\'a}ndez
  Santisteban}, {Gallo}, {Atri}, {Plotkin}, {Maccarone}, {Sivakoff}, {Miller},
  {Reynolds}, {Russell}, {Maitra}, {Heinke}, {Armas Padilla}, \&
  {Shaw}}]{van-den-eijnden_et_al_2021}
{van den Eijnden}, J., {Degenaar}, N., {Russell}, T.~D., {et~al.} 2021, \mnras,
  507, 3899

\bibitem[{Virtanen {et~al.}(2020)Virtanen, Gommers, Oliphant, Haberland, Reddy,
  Cournapeau, Burovski, Peterson, Weckesser, Bright, {van der Walt}, Brett,
  Wilson, Millman, Mayorov, Nelson, Jones, Kern, Larson, Carey, Polat, Feng,
  Moore, {VanderPlas}, Laxalde, Perktold, Cimrman, Henriksen, Quintero, Harris,
  Archibald, Ribeiro, Pedregosa, {van Mulbregt}, \& {SciPy 1.0
  Contributors}}]{SciPy}
Virtanen, P., Gommers, R., Oliphant, T.~E., {et~al.} 2020, Nature Methods, 17,
  261

\bibitem[{{Wendker}(1995)}]{Wendker_1995}
{Wendker}, H.~J. 1995, \aaps, 109, 177

\bibitem[{{Wendker}(2015)}]{Wendker_2015}
{Wendker}, H.~J. 2015, VizieR Online Data Catalog, VIII/99

\bibitem[{{Williams} {et~al.}(2010){Williams}, {Gies}, {Matson}, {Touhami},
  {Grundstrom}, {Huang}, \& {McSwain}}]{Williams_et_al_2010}
{Williams}, S.~J., {Gies}, D.~R., {Matson}, R.~A., {et~al.} 2010, \apjl, 723,
  L93

\bibitem[{{Zamanov} {et~al.}(2022){Zamanov}, {Stoyanov}, {Marchev}, {Tomov},
  {Wolter}, {Bode}, {Nikolov}, {Stefanov}, {Kurtenkov}, \&
  {Latev}}]{Zamanov_et_al_2022}
{Zamanov}, R.~K., {Stoyanov}, K.~A., {Marchev}, D., {et~al.} 2022,
  Astronomische Nachrichten, 343, e24019

\bibitem[{{Zamanov} {et~al.}(2016){Zamanov}, {Stoyanov}, {Mart{\'\i}}, {Latev},
  {Nikolov}, {Bode}, \& {Luque-Escamilla}}]{Zamanov_et_al_2016}
{Zamanov}, R.~K., {Stoyanov}, K.~A., {Mart{\'\i}}, J., {et~al.} 2016, \aap,
  593, A97

\bibitem[{{Zucker}(2003)}]{Zucker_2003}
{Zucker}, S. 2003, \mnras, 342, 1291

\end{thebibliography}

\newpage~\newpage
\titleformat{\section}{\large\bfseries}{\appendixname~\thesection .}{0.5em}{}
\begin{appendices}
\renewcommand{\thefigure}{A\arabic{figure}}
\setcounter{figure}{0}
\renewcommand{\thetable}{A\arabic{table}}
\setcounter{table}{0}

\section{RV measurements}
Table \ref{table_appendix_RVs} lists the measured RVs for the HERMES spectra.

\begin{minipage}{\textwidth}
    \begin{center}
       % \begin{table}[!hbp]
% \renewcommand{\arraystretch}{1.1}
    % \centering
    \captionof{table}{Measured RVs for the given lines in each HERMES spectrum.}
    \begin{tabular}{ ccrrrr }
     \hline
     \hline
     phase & bjd & $v_{\text{He {\sc ii} $\lambda$4686}}$ [km s$^{-1}$] & $v_{\text{He {\sc i} lines}}$ [km s$^{-1}$] & $v_{\text{H$\beta$}}$ [km s$^{-1}$] & $v_{\text{Fe {\sc ii} $\lambda$4583}}$ [km s$^{-1}$] \\ 
     \hline
      0.093 & 2459462.614 & $45.1 \pm 4.2$ & $-1.0 \pm 3.6$ & $-17.6 \pm 1.7$ & $-26.1 \pm 2.6$\\
      0.165 & 2459053.691 & $-1.0 \pm 2.3$ & $-7.0 \pm 2.4$ & $-0.1 \pm 0.9$ & $-17.6 \pm 1.8$\\
      0.233 & 2459057.692 & $-11.1 \pm 3.0$ & $-5.1 \pm 2.5$ & $-2.6 \pm 1.3$ & $-27.2 \pm 2.1$\\
      0.281 & 2459119.538 & $-4.4 \pm 3.9$ & $-6.6 \pm 2.5$ & $-2.8 \pm 1.2$ & $-21.1 \pm 1.7$\\
      0.283 & 2459060.618 & 0.2 $\pm$ 2.9 & $-6.3 \pm 2.3$ & $-3.8 \pm 1.3$ & $-24.3 \pm 1.5$\\
      0.286 & 2456227.456 & 0.0 $\pm$ 3.2 & 0.0 $\pm$ 2.3 & 0.0 $\pm$ 1.6 & 0.0 $\pm$ 2.9\\
      0.399 & 2456647.321 & 47.3 $\pm$ 3.7 & $-2.8 \pm 3.6$ & $-9.6 \pm 1.5$ & $-32.5 \pm 2.2$\\
      0.467 & 2459130.500 & 78.2 $\pm$ 3.3 & $-11.4 \pm 2.8$ & $-14.9 \pm 1.0$ & $-28.6 \pm 2.1$\\
      0.482 & 2459426.569 & 89.1 $\pm$ 3.3 & $-8.8 \pm 3.0$ & $-16.2 \pm 1.3$ & $-29.1 \pm 2.2$\\
      0.546 & 2457659.462 & 139.9 $\pm$ 3.3 & $-17.9 \pm 3.1$ & $-3.3 \pm 2.3$ & $-26.5 \pm 2.4$\\
      0.571 & 2459077.606 & 109.6 $\pm$ 2.1 & $-24.2 \pm 2.5$ & $-20.5 \pm 1.4$ & $-40.5 \pm 2.1$\\
      0.612 & 2457663.402 & 182.6 $\pm$ 4.2 & $-11.5 \pm 3.5$ & $-4.0 \pm 1.8$ & $-25.9 \pm 2.9$\\
      0.638 & 2459081.566 & 134.4 $\pm$ 2.5 & $-27.1 \pm 2.3$ & $-21.4 \pm 1.7$ & $-36.6 \pm 1.7$\\
      0.685 & 2459733.687 & 178.8 $\pm$ 3.4 & $-31.4 \pm 3.2$ & $-20.5 \pm 1.7$ & $-38.6 \pm 2.3$\\
      0.773 & 2459384.721 & $181.2 \pm 2.4$ & $-23.6 \pm 3.6$ & $-17.5 \pm 1.9$ & $-60.6 \pm 2.6$\\
      0.857 & 2459094.506 & $158.6 \pm 1.8$ & $-24.0 \pm 2.3$ & $-19.4 \pm 1.2$ & $-41.1 \pm 1.6$\\
      0.890 & 2459096.467 & $146.0 \pm 1.8$ & $-20.2 \pm 2.3$ & $-22.3 \pm 1.1$ & $-47.0 \pm 2.1$\\
      0.991 & 2459102.410 & $90.6 \pm 2.1$ & $-16.7 \pm 2.8$ & $-22.8 \pm 0.9$ & $-35.1 \pm 1.6$\\
      \hline
    \end{tabular}
         \flushleft
    \begin{tablenotes}
      \small
      \item \textbf{Notes.} Listed RVs are calibrated to have RV = 0 km s$^{-1}$ for the spectrum taken at bjd = 2456227.456.
    \end{tablenotes}
\label{table_appendix_RVs}
% \end{table}

    \end{center}
\end{minipage}

%\newpage%~\newpage
\section{Spectral features}
Figure \ref{fig_CaIItriplet} shows several regions of the spectra of MWC 656. Three spectra at phases 0.23, 0.77, and 0.98 are shown.
\begin{figure*}[!hb]
    \centering
    \begin{subfigure}{\linewidth}
        \includegraphics[width = \textwidth]{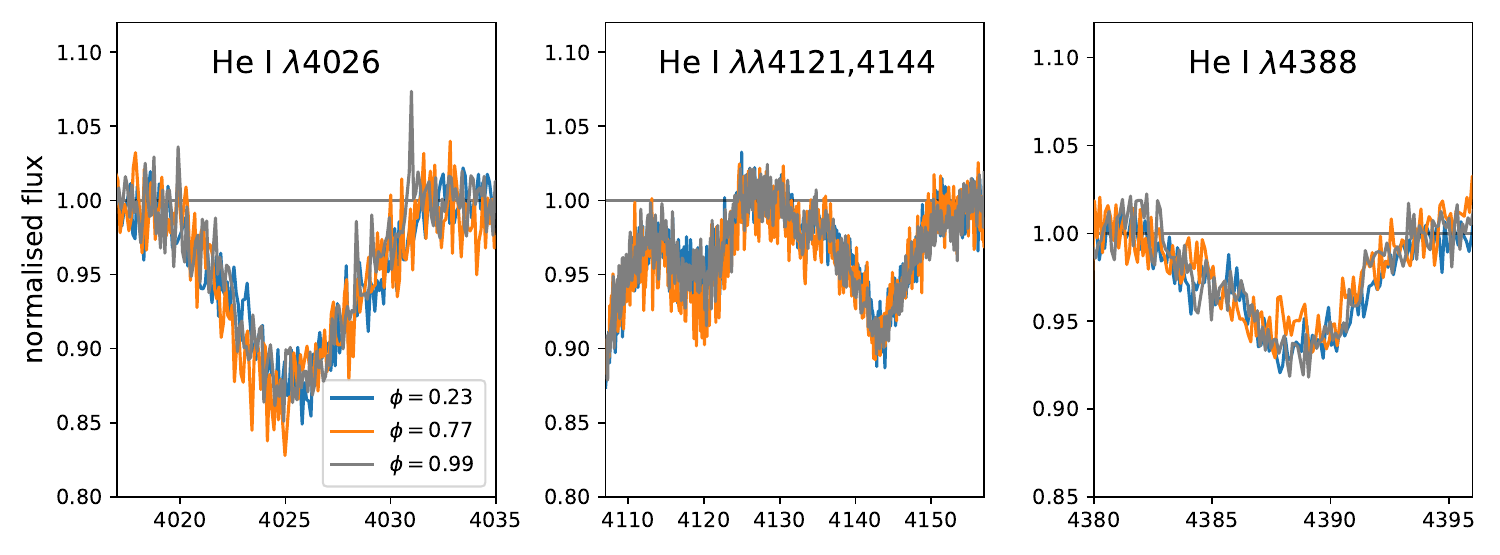}
    \end{subfigure}
    \begin{subfigure}{\linewidth}
        \includegraphics[width = \textwidth]{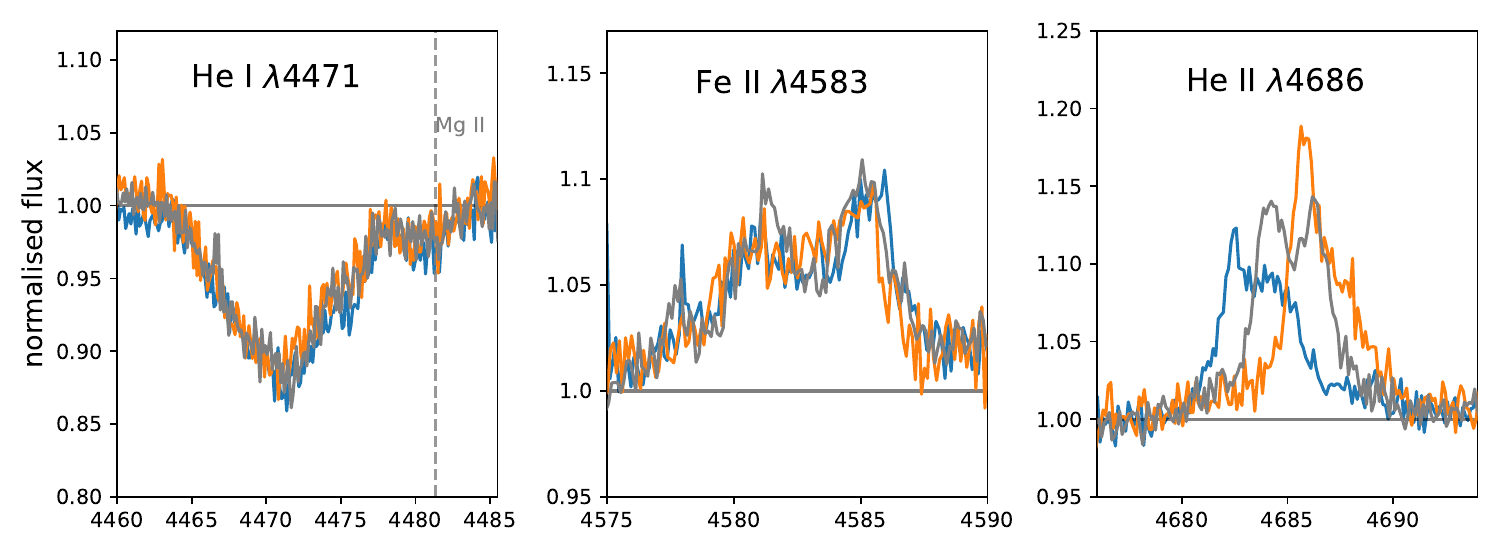}
    \end{subfigure}
    \begin{subfigure}{\linewidth}
        \includegraphics[width = \textwidth]{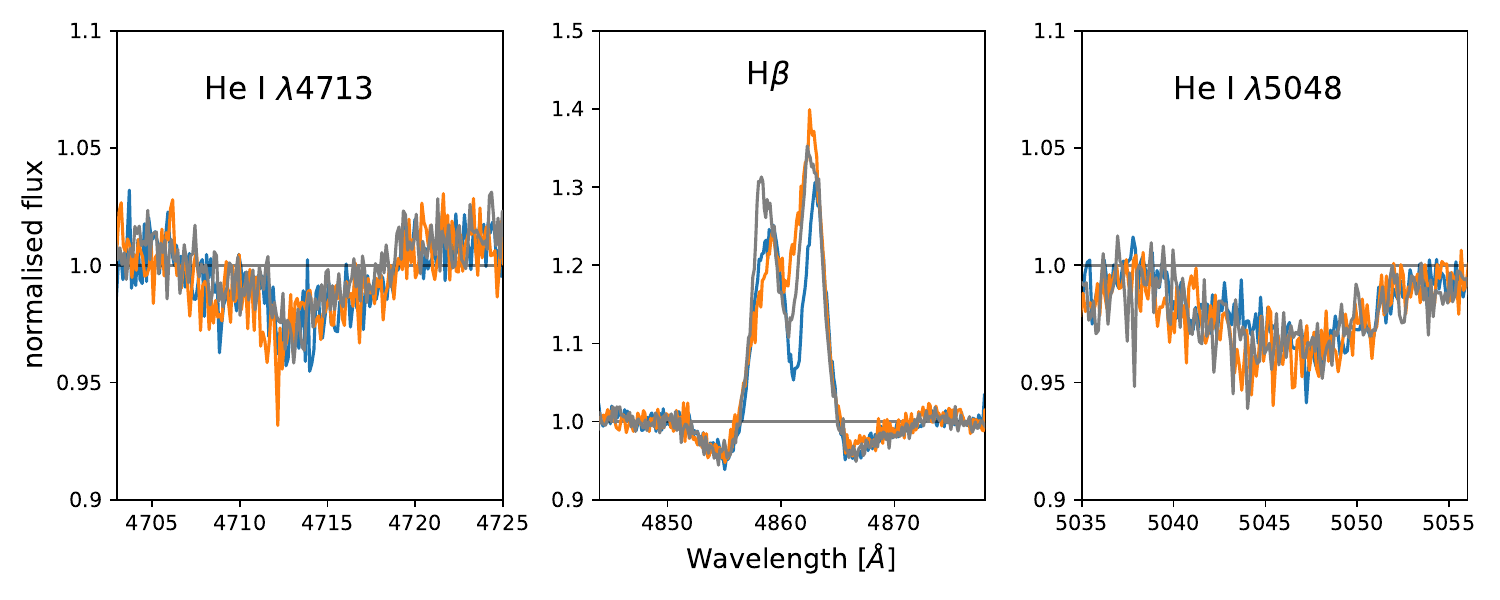}
    \end{subfigure}
    \caption{HERMES spectra of MWC 656 for phases corresponding to 0.23 (blue), 0.77 (orange), and 0.98 (grey). Each panel zooms in on a specific spectral line region, indicated in the individual panels. The left-most absorption feature in the panel of He {\sc i} $\lambda\lambda$4121,4144 is part of the red wing of H$\delta$. %Top: from left to right H$\alpha$, H$\beta$, He {\sc ii} $\lambda$4686. Bottom: part of the Paschen series and the Ca {\sc ii} triplet. 
    The grey spectrum shows a clearly double peaked He {\sc ii}.
    }
    \label{fig_CaIItriplet}
\end{figure*}
\begin{figure*}[!hb]
    \ContinuedFloat
    \begin{subfigure}{\linewidth}
        \includegraphics[width = \textwidth]{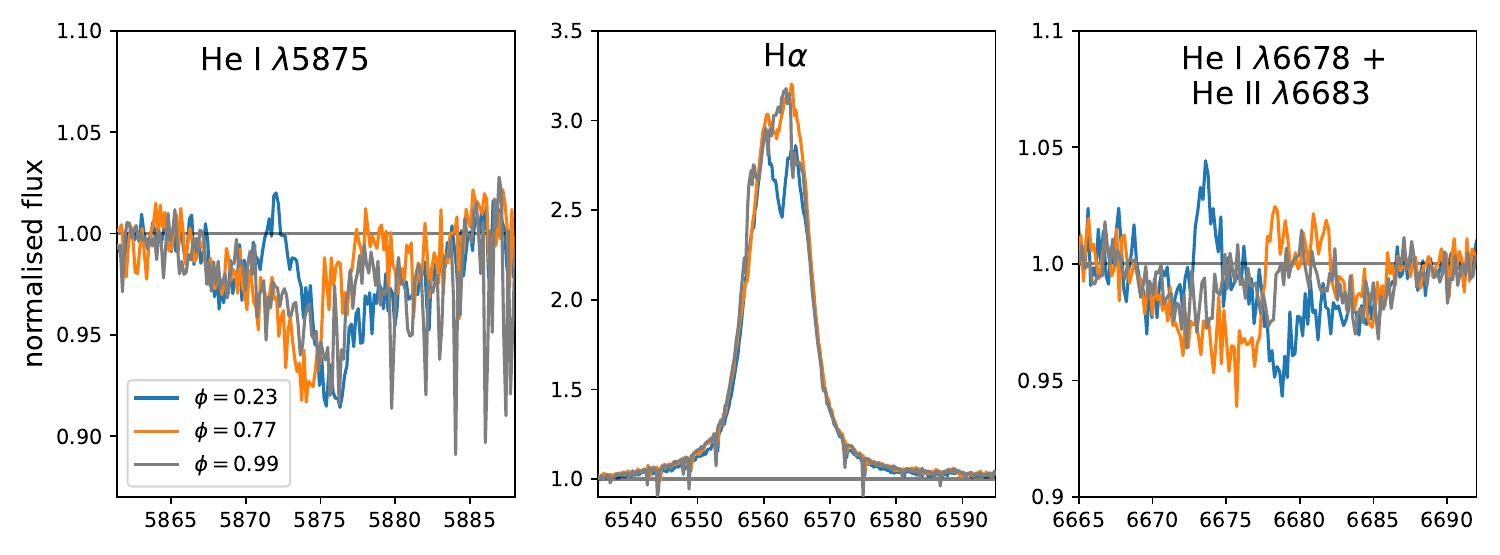}
    \end{subfigure}
    \begin{subfigure}{\linewidth}
        \includegraphics[width = \textwidth]{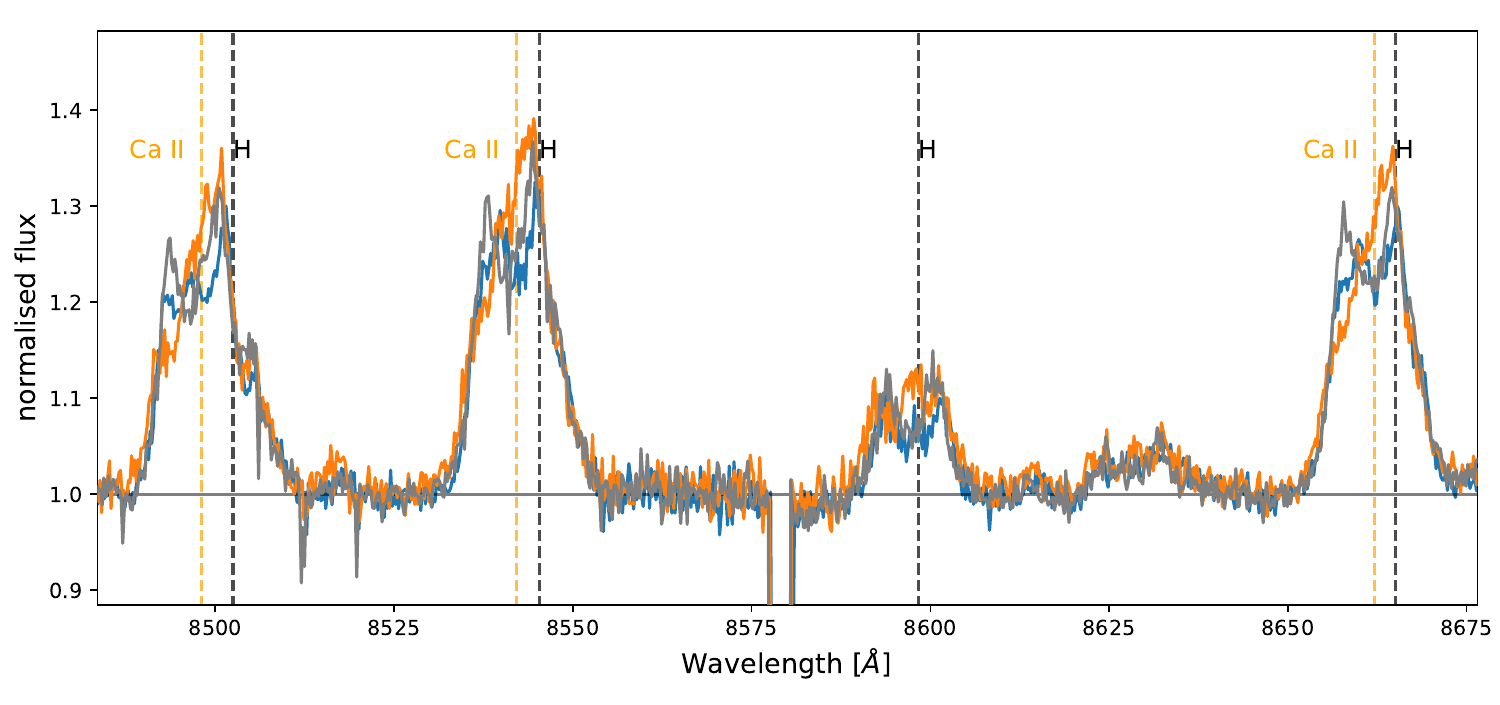}
    \end{subfigure}
    \caption{Continued. The panels with He {\sc i} $\lambda$5875 and He\,{\sc i} $\lambda$6678 + He {\sc ii} $\lambda$ 6683 show clear blending with another emission component. The panel showing He {\sc i} $\lambda$5875 is also contaminated by telluric lines. The bottom panel shows part of the Paschen series and the Ca {\sc ii} triplet.}

\end{figure*}

\newpage~\newpage~\newpage~\newpage~\newpage~\newpage
\section{Dynamical spectra}
\begin{figure*}[!hb]
    \centering
    \begin{subfigure}{\linewidth}
        \includegraphics[width = \textwidth]{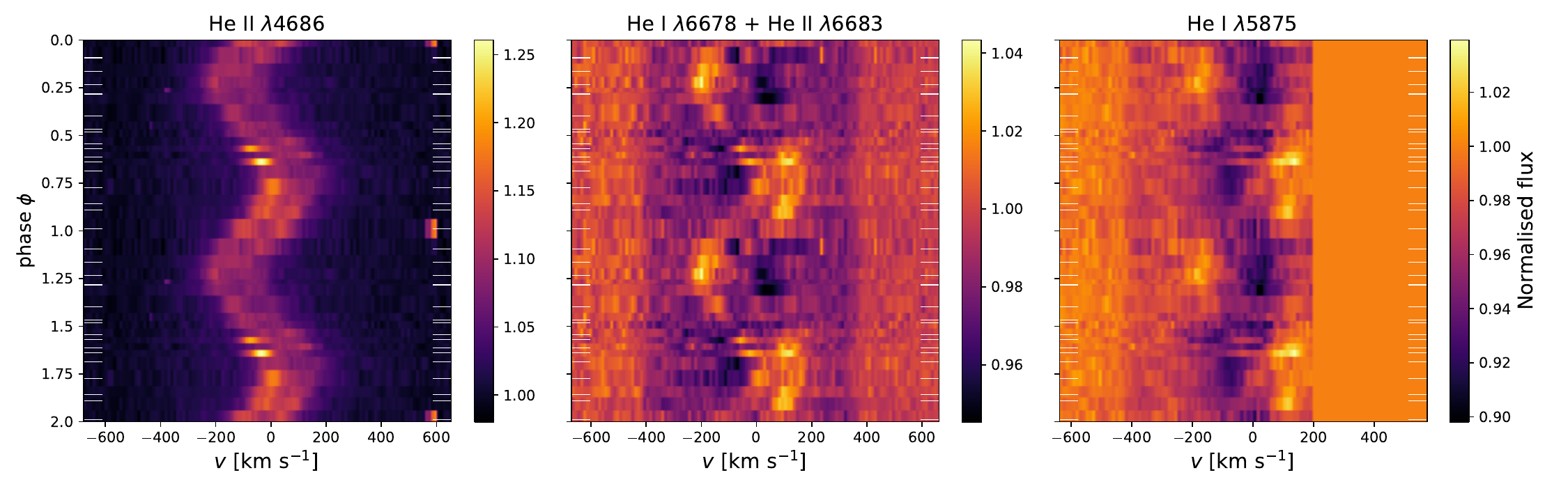}
    \end{subfigure}
    % \vspace{-0.5in}
    \begin{subfigure}{\linewidth}
        \includegraphics[width = \textwidth]{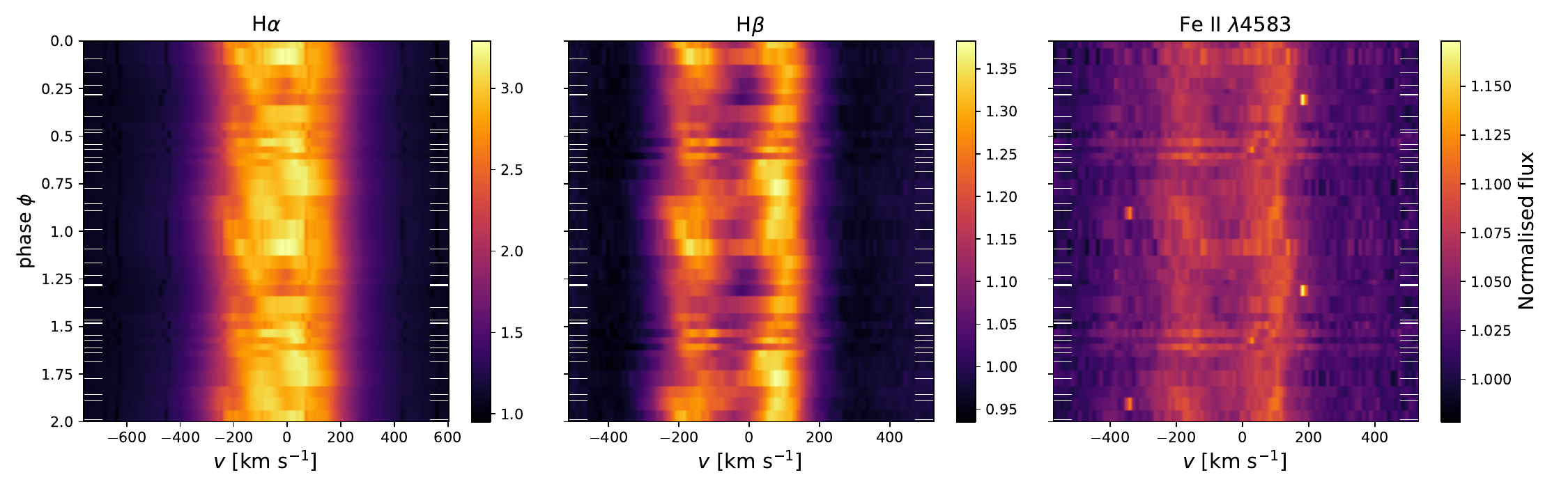}
    \end{subfigure}
    \caption{Dynamical spectra of MWC 656. Phases are shown for two orbital periods. White horizontal bars indicate the phases at which HERMES spectra are taken. Top, from left to right: He {\sc ii} $\lambda$4686, He {\sc i} $\lambda$6678 + He {\sc ii} $\lambda$6683, He {\sc i} $\lambda$5875. Bottom, from left to right: H$\alpha$, H$\beta$, Fe {\sc ii} $\lambda$4583. For He {\sc i} $\lambda$5875, part of the spectra where $v>200$\,km\,s$^{-1}$ is put equal to one because of telluric contamination and contamination from the interstellar Na {\sc i} line. The top-middle (He {\sc i} $\lambda$6678 + He {\sc ii} $\lambda$6683) and -right (He {\sc i} $\lambda$5875) panels trace similar motion in emission as He {\sc ii} $\lambda$4686 on top of the absorption coming from the Be star.}
    \label{fig_dynamical}
\end{figure*}

Figure \ref{fig_dynamical} shows the dynamical spectra of MWC 656 for different spectral lines.

\newpage~\newpage
\section{Testing the limits of spectral disentangling}\label{appendix_spectral_disentangling}
Spectral disentangling is a useful tool for the detection of very faint companions and for putting constraints on the light ratio \citep[e.g.][]{Bodensteiner_et_al_2020,Shenar_et_al_2020,Shenar_et_al_2022}. However, if companions are too faint compared to the S/N of the spectroscopic data, even disentangling cannot detect them. In order to test to which light ratios disentangling could detect a helium companion in MWC 656, we performed simulations on mock spectra.

For the Be star, the B-star spectrum was simulated with a TLUSTY \citep[][]{TLUSTY} model spectrum of a 19\,000\,K star with $\log g = 4.00$\,dex and $v \sin i = 330$\,km\,s$^{-1}$. For the stripped star, we created models using the Potsdam Wolf-Rayet (PoWR) code \citep[][]{PoWR} that agree with the expected parameters for the stripped star. The lowest possible mass of 1\,$M_{\odot}$ yields parameters of $T_{\text{eff}} \sim 45\,000$\,K and $R \sim 0.4\,R_{\odot}$, while the upper mass limit of $\sim$ 2.5\,$M_{\odot}$ results in a stripped star with $T_{\text{eff}} \sim 60\,000$\,K and $R \sim 0.6\,R_{\odot}$ \citep[][]{Gotberg_et_al_2018}. Combining this with the parameters of the Be star, we estimate an optical contribution of the stripped star between $\lesssim$\,1\% and $\sim$\,3\% for the lowest and highest mass, respectively.

As explained in Sect. \ref{sec_disentangling_results}, we do not disentangle on the Balmer lines. Based on figure 5 in \citet{Gotberg_et_al_2018}, all stripped star models in the derived mass range seem to show absorption in He {\sc i}(+{\sc ii}) $\lambda$4026 and He {\sc ii} $\lambda$4542. The lower-mass models also show absorption in He {\sc i} $\lambda$4472. Therefore, we focused disentangling on these three lines.

\begin{figure*}[!hb]
    \centering
    \includegraphics[width = \textwidth]{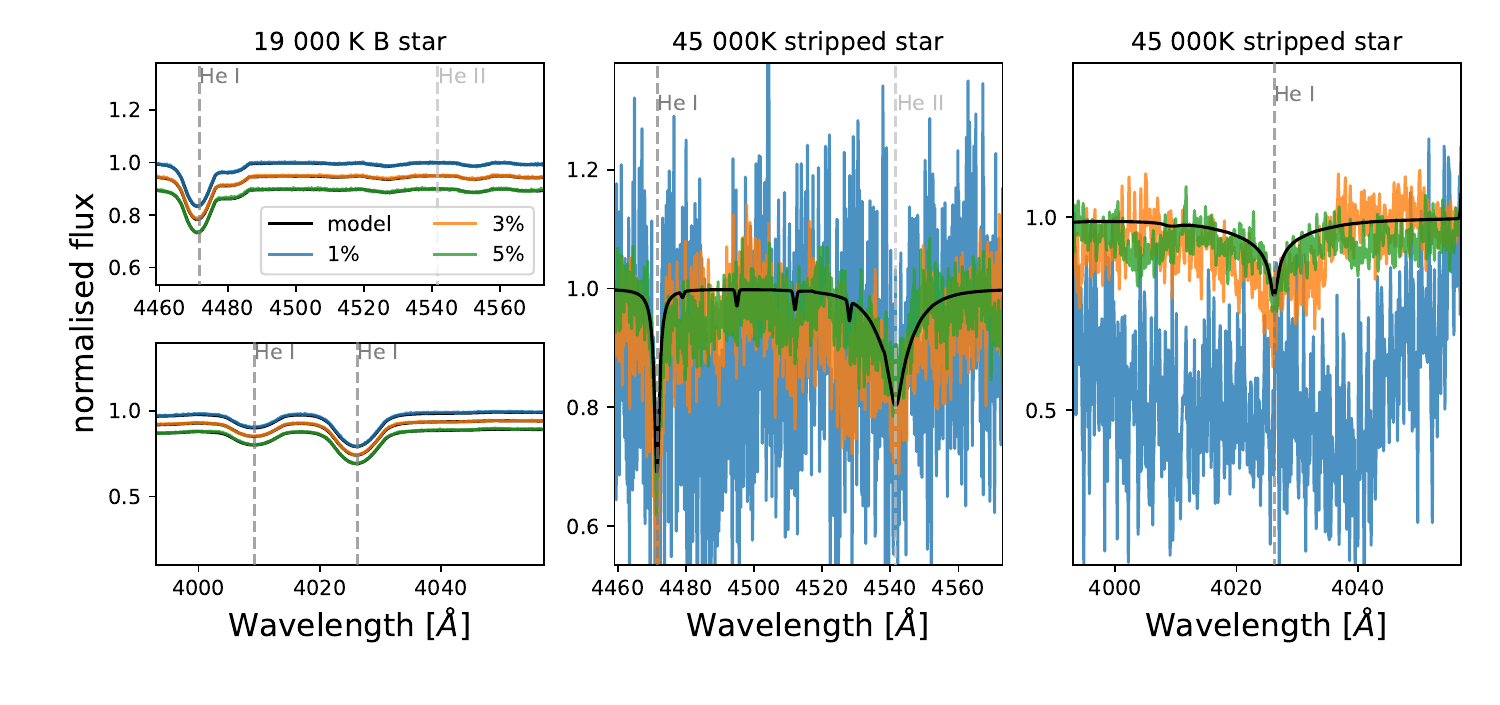}
    \caption{Disentangled simulated spectra for a 45\,000\,K stripped star with different contributions. Black are the models, blue are the disentangled spectra given a 1\% contributed flux, orange for a 3\% contributed flux, and green for a 5\% contributed flux. The model in the left panels is for a B star of 19\,000\,K and $v\sin i = 330$\,km s$^{-1}$. The model in the middle and right panels is for a stripped star of 45\,000\,K and $v\sin i = 30$\,km s$^{-1}$.}
    \label{fig_45000K}
\end{figure*}

For a stripped $1M_{\odot}$ star with $T_{\text{eff}} = 45$\,kK, we created mock spectra with a flux contribution for the stripped star of 1\%, 3\%, and 5\%. A companion with a 3\% light contribution would be detected using He {\sc i} $\lambda$4471 and He {\sc i}(+{\sc ii}) $\lambda$4026. However, the expected 1\% contribution seems to be very challenging, as, for example, He {\sc i}(+{\sc ii}) $\lambda$4026 is not detected (see Fig. \ref{fig_45000K}).

\begin{figure*}[!hb]
    \centering
    \includegraphics[width = \textwidth]{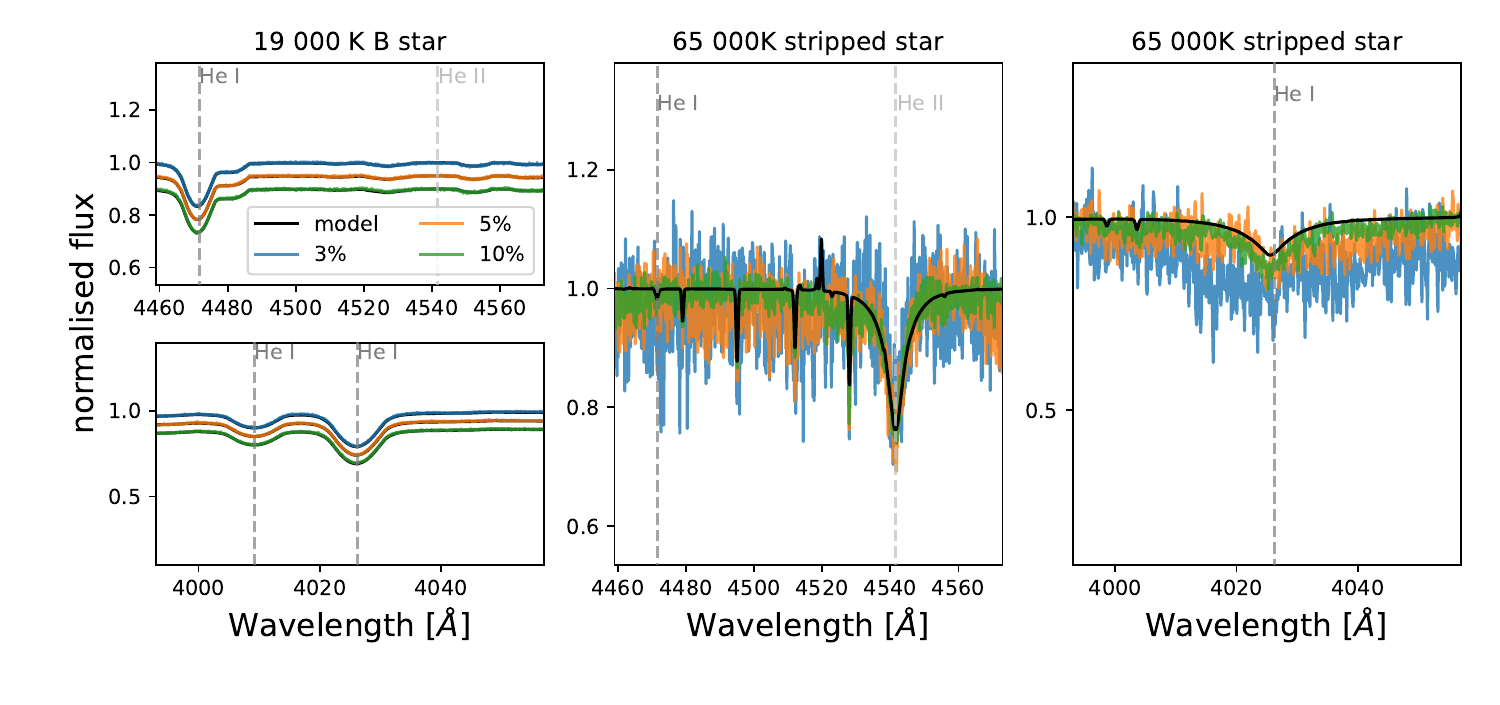}
    \caption{Same as Fig. \ref{fig_45000K}, but for a 65\,000\,K stripped star with a 3\% (blue), 5\% (orange), and 10\% (green) contributed flux.}
    \label{fig_65000K}
\end{figure*}

For a $2.5 M_{\odot}$ stripped star of $T_{\text{eff}} = 60$\,kK , we created mock spectra with a flux contribution for the stripped star of 3\%, 5\%, and 10\%. Disentangling seems possible down to a flux contribution of 5\%. However, the expected 3\% flux contribution proves very challenging. While the He {\sc ii} $\lambda$4542 seems detectable, again the He {\sc i}(+{\sc ii}) $\lambda$4026 is not detected (see Fig. \ref{fig_65000K}). Thus, based on simulations using mock spectra, we do not expect to be able to detect a stripped star companion using spectral disentangling.


\end{appendices}

\end{document}